\documentclass[preprint,12pt]{elsarticle}
\usepackage{float}

\usepackage{amssymb}
\usepackage{amsmath}
\usepackage{graphicx}
\usepackage{subfig}
\usepackage{booktabs}

\journal{Nuclear Engineering and Technology} 

\begin{document}

\begin{frontmatter}



\title{Applicability of a Thermodynamic Model to Rich-Hydrogen Flammability Assessment in SMR Containments}


\author[a]{Sihyeong Yu} 
\author[b]{Dosu Park}
\author[a]{Jeesuk Shin}
\author[a]{Joongoo Jeon\corref{cor1}}

\cortext[cor1]{Corresponding author: jgjeon41@postech.ac.kr}

\affiliation[a]{organization={Division of Advanced Nuclear Engineering, Pohang University of Science and Technology (POSTECH)},
            addressline={77 Cheongam-ro, Nam-gu}, 
            city={Pohang},
            postcode={37673}, 
            country={Republic of Korea}} 

\affiliation[b]{organization={Department of Quantum System Engineering, Jeonbuk National University},
            addressline={567 Baekje-daero}, 
            city={Jeonju},
            postcode={54896}, 
            country={Republic of Korea}}

\begin{abstract}
Hydrogen combustion remains a critical safety concern during severe accidents in small modular reactors (SMRs). The compact containment can develop rich-hydrogen atmospheres through zirconium oxidation while oxygen gradually accumulates via radiolysis. Conventional flammability assessment methodologies, however, have been validated and applied only for lean-hydrogen atmospheres. This study establishes a thermodynamic flammability assessment method for rich-hydrogen conditions using the Calculated Adiabatic Flame Temperature (CAFT). In rich mixtures the deficient reactant becomes oxygen and the Lewis number exceeds unity, matching the hydrocarbon regime where CAFT is established. The CAFT was evaluated at 21 points along the measured upper flammability limit boundary of the U.S. NRC FITS database. The limit temperature remains quasi-constant within 17\% across steam fractions of 5--50\%. Its minimum value of 1112.5 K was adopted as a conservative threshold. The criterion was applied to the NuScale severe accident scenario LCC-05T-03. A transient flammable window is identified during the 44--48 hour interval, when 8--10\% oxygen coexists with vigorous hydrogen generation and low steam fractions. These results show that the oxygen criterion alone cannot resolve transient flammability windows. The CAFT criterion enables more comprehensive flammability assessment for SMR containments.

\end{abstract}


\end{frontmatter}



\section{Introduction}
\label{sec1}
Hydrogen combustion risk constitutes a critical safety concern during severe accidents in the nuclear industry \cite{nishimura2015current}. Large quantities of hydrogen are generated through the exothermic oxidation reaction between zirconium cladding and high-temperature steam, and subsequently released into the containment atmosphere \cite{cronenberg1990vessel}. If the local mixture composition lies within the flammable range and an ignition source is present, combustion can occur. Flame acceleration (FA) and deflagration-to-detonation transition (DDT) may then follow depending on the mixture and geometric conditions \cite{hong2015research, kim2018systematic}. The Three Mile Island Unit 2 (TMI-2) accident in 1979 and the Fukushima Daiichi accident in 2011 demonstrated the severity of hydrogen combustion risks in practical terms \cite{nishimura2015current, camp1983light}.

In particular, hydrogen combustion in severe accident scenarios of small modular reactors (SMRs) exhibits characteristics qualitatively different from those of conventional large-scale nuclear power plants. The NuScale SMR adopts a vacuum containment vessel design, resulting in a very low initial oxygen concentration. However, the limited containment volume can lead to the formation of rich-hydrogen conditions during zirconium oxidation \cite{weber2020severe, a2019_assessment}. A similar vacuum containment concept is used in other integral PWR-type SMRs. The Korean i-SMR maintains its containment vessel under vacuum with a containment atmosphere control system during normal operation~\cite{lim2025design}. The Westinghouse SMR also uses a compact partial-vacuum containment derived from the AP1000 passive safety architecture~\cite{smith2012westinghouse, alamgir2016design}. These conditions are fundamentally different from the lean-hydrogen conditions that have been the primary focus of conventional large-scale plants, necessitating a dedicated methodology for evaluating hydrogen combustion potential.

Current hydrogen flammability assessment in nuclear severe accident analysis relies predominantly on empirical methods, such as the Shapiro diagram and the flammability correlations derived from the FITS experiments, which are adopted in severe accident codes like MELCOR and MAAP \cite{shapiro1957hydrogen, bentaib2015overview, kumar1985flammability, plys1993hydrogen, marshall1986hydrogen, jeon2020recent}. These experimental maps do cover compositions rich in hydrogen, including the upper flammability boundary. Their limitation lies elsewhere. The underlying measurements were performed under fixed reference conditions, namely an air background near atmospheric pressure and a prescribed initial temperature. The resulting maps return a binary verdict for a queried composition and carry no information on the thermodynamic state of the mixture. Severe accident atmospheres in SMR containments do not satisfy these reference conditions, because the background gas deviates from air due to oxygen depletion and the pressure, temperature, and composition evolve rapidly \cite{weber2020severe, a2019_assessment}. A lookup in composition coordinates therefore offers neither a physical basis for extrapolation nor a quantitative measure of the margin to the flammability limit.

As an analytical alternative to empirical correlations, the Calculated Adiabatic Flame Temperature (CAFT) model has been widely used for flammability assessment based on a systematic thermodynamic energy balance, demonstrating reasonable accuracy for predicting both the LFL and UFL of hydrocarbon fuels \cite{vidal2006evaluation, mavcek1979flammability, mashuga1999flammability, wu2018new}. For hydrogen, the conventional CAFT model has been shown to exhibit reduced accuracy under specific conditions involving radiative heat loss, and a non-adiabatic extension has been developed to address this limitation on the lean-hydrogen side \cite{jeon2019flammability}. Although CAFT-based predictions of hydrogen UFL have been reported for dry $H_2$-air mixtures \cite{terpstra2012flammability, liaw2016model}, their applicability under SMR severe accident conditions remains unexamined. In these accidents elevated temperature, elevated pressure, and substantial steam dilution coexist in the rich-hydrogen regime.

A thermal-diffusive consideration suggests that the rich-hydrogen regime is, in fact, favorable for the CAFT approach. In lean-hydrogen mixtures, the deficient reactant $H_2$ has a Lewis number substantially below unity, and the resulting strong preferential diffusion produces flame behavior distinct from that of conventional hydrocarbon flames. In fuel-rich hydrogen mixtures, by contrast, the deficient reactant shifts to $O_2$ and the effective Lewis number exceeds unity ($Le>1$) \cite{burbano2011laminar}. This transition brings the thermal-diffusive behavior of rich hydrogen flames close to that of hydrocarbon flames. CAFT-based UFL prediction has been demonstrated to perform reliably in that regime \cite{wu2018new}. This physical correspondence motivates the hypothesis that the conventional CAFT model provides a reasonable thermodynamic basis for flammability assessment under rich-hydrogen conditions. To date, however, this hypothesis has not been empirically tested, leaving an open question of substantial relevance to SMR severe accident analysis.

\begin{figure}[!t]
    \centerline{\includegraphics[width=\columnwidth]{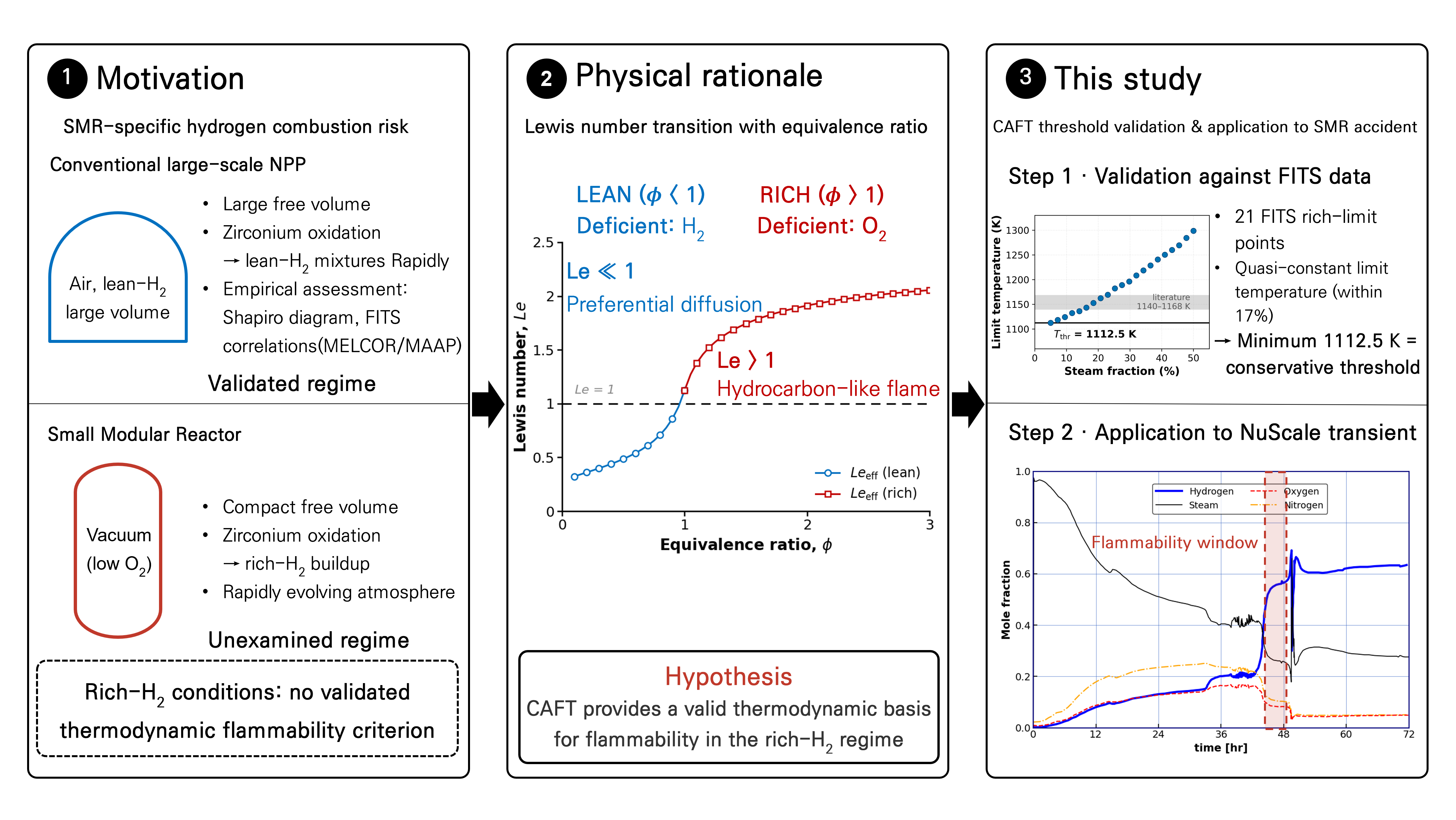}}
    \caption{Overview of the present study. (1)~Motivation: empirical flammability criteria established at fixed reference conditions provide no thermodynamic basis for the rich-$\mathrm{H_2}$ atmospheres of SMR severe accidents. (2)~Physical rationale: the effective Lewis number of $\mathrm{H_2}$--air mixtures transitions from $Le \ll 1$ under lean conditions to $Le > 1$ under rich conditions, where the thermal-diffusive behavior approaches that of hydrocarbon flames.Effective Lewis numbers computed in this work with mixture-averaged transport (Section~2.3). (3)~Approach: the CAFT is evaluated along the measured UFL boundary of the FITS database, and its minimum value of 1112.5 K defines a conservative flammability threshold. The criterion is then applied to a NuScale severe accident transient to identify flammability windows not captured by the conventional $\mathrm{O_2}$ criterion.}
    \label{fig:overview}
\end{figure}

The present study addresses this gap through two coupled objectives (Fig.~\ref{fig:overview}). First, the CAFT is evaluated along the measured UFL boundary of the FITS database to test the constancy of the limit temperature. The minimum limit value defines a conservative flammability threshold, and its classification performance and applicable envelope are quantified. This constitutes the first validation of CAFT for rich-hydrogen mixtures that combines a mechanism-based applicability argument with quantitative comparison against experiments. Second, the validated criterion is applied to a NuScale SMR severe accident scenario to assess flammability along the transient evolution of the containment atmosphere. This analysis identifies an early transient flammability window that the conventional oxygen concentration criterion does not resolve.

\section{Background and Literature Review}
\subsection{Hydrogen Risk in Nuclear Severe Accidents}
\label{sec:hydrogen_risk}

    Hydrogen generation during nuclear severe accidents can be broadly divided into two phases. During the in-vessel phase, the zirconium fuel cladding undergoes an exothermic oxidation reaction with high-temperature steam, producing hydrogen gas ($Zr + 2H_2O \to ZrO_2 + 2H_2$) \cite{cronenberg1990vessel}. During the ex-vessel phase, the molten corium–concrete interaction (MCCI) generates hydrogen along with carbon monoxide \cite{fineschi2001mitigation}, which is also a combustible gas \cite{kim2020improved, cohen1992flammability}. The behavior of hydrogen within the containment is governed by complex phenomena including natural convection, steam condensation, and the operation of passive autocatalytic recombiners (PARs). These general features, however, apply differently to the NuScale design analyzed in this study. The reactor vessel sits inside a compact steel containment vessel that is submerged in the reactor pool. Condensate collecting in the containment vessel submerges the reactor vessel lower head and provides a conduction path for decay heat to the reactor pool, so core debris is retained in-vessel. This retention is supported by an independent NRC assessment \cite{nrc2018ivr}, and the confirmatory MELCOR analysis of the reference scenario predicts that vessel breach is prevented even without ECCS actuation \cite{a2019_assessment}. Because there is no concrete inside the steel containment vessel, ex-vessel MCCI is effectively precluded and carbon monoxide has no significant generation pathway. The present study therefore considers $H_2$–air–steam mixtures only. 
The design variant analyzed in the ERI report employs no PARs, and combustible gas control relies on the evacuated containment, which keeps the atmosphere depleted of oxygen until radiolysis gradually restores it \cite{a2019_assessment}. The subsequently approved US460 standard design adds a passive autocatalytic recombiner in each module to keep the containment oxygen below 4\% \cite{a2025_safety}. The present analysis therefore applies to the design variant documented in the ERI report. Hydrogen behavior is then governed by condensation on the pool-cooled containment wall and by rapid composition transients in the compact free volume.
Research on hydrogen flammability limits was greatly intensified following the TMI-2 accident. The Fully Instrumented Test Site (FITS) experiments conducted by Sandia National Laboratories represent a benchmark experimental program for measuring the flammability limits of hydrogen–air–steam mixtures, performed at an initial temperature of approximately 110~$^\circ$C \cite{marshall1986hydrogen}. In the FITS experiments, flammability was determined based on the measured pressure rise, and the variation of flammability limits with steam concentration was systematically investigated. Current severe accident analysis codes, including MAAP and MELCOR, continue to rely on such experimental results for hydrogen combustion risk assessment \cite{jeon2020recent}.
Conventional large pressurized water reactor (PWR) containment buildings are designed with large free volumes (approximately $5\sim 8 \times 10^4 ~ m^3$), providing substantial margin for hydrogen dilution \cite{camp1983light}. In contrast, small modular reactors such as the NuScale SMR feature significantly smaller containment volumes, which can lead to elevated local hydrogen concentrations even for the same amount of hydrogen generation \cite{weber2020severe}. Furthermore, the design characteristics of SMR containments may maintain high steam partial pressures, underscoring the need for flammability assessment methodologies applicable to rich-hydrogen, high-steam conditions \cite{weber2020severe}.

\subsection{Flammability Limit Prediction Methodologies}

Flammability limit prediction methodologies can be broadly classified into experimental, numerical, and analytical approaches.
Experimental methods have been employed since the 19th century, and a standardized test procedure was established by Coward and Jones (Bureau of Mines standard: open tube with 5 cm diameter and 1.5 m length) \cite{coward1952limits}. Subsequent studies have measured flammability limits under various conditions of temperature, pressure, and diluent composition. While the LFL of hydrogen in air at ambient conditions is well established at approximately 4.1 vol\%, it is impractical to experimentally determine the flammability limits for all possible combinations of diluent type, concentration, initial temperature, and pressure.

Numerical methods have advanced from one-dimensional planar flame analyses to multi-dimensional stretched flame simulations, driven by improvements in computational capabilities. Numerical simulations coupled with detailed chemical kinetic mechanisms (e.g., the San Diego mechanism) have enabled microscopic analysis of flame structure \cite{saxena2006testing, sanchez2014recent}. Zhou et al. employed the stabilized flame method to simulate the extinction processes of hydrogen, methane, and hydrogen–methane blend flames in two dimensions, and attributed the differences in extinction mechanisms among fuel types to Lewis number effects \cite{zhou2018experimental, zhou2018effect}. However, numerical approaches are subject to inherent uncertainties in chemical kinetics, binary diffusion coefficients, and radiation models, and the computational cost for diverse mixture conditions remains substantial. Representative numerical studies on the flammability limit are summarized in Table~\ref{tab:numerical_studies}.

Analytical methods predict flammability limits through macroscopic energy balance of the flame. The CAFT model is a representative example \cite{egerton1953limits, zabetakis1965flammability}, and thermal loss theories by Spalding and Mayer, as well as flame stretch theories by Buckmaster et al., have been developed \cite{spalding1957theory, mayer1957theory, buckmaster1982flammability}. 
These analytical approaches offer the advantage of rapid flammability limit prediction without detailed chemical kinetic calculations.

Their accuracy, however, is not uniform across fuels and regimes. It depends on whether the limit phenomenon is governed by the flame energy budget that these models resolve. When the limit is set by the thermal balance of a smooth flame, as in hydrocarbon fuels with Lewis numbers close to unity, energy balance methods perform well \cite{vidal2006evaluation, mavcek1979flammability, mashuga1999flammability}. When transport processes neglected by these models dominate, as in lean hydrogen flames with $Le \ll 1$, their accuracy deteriorates \cite{jeon2019flammability, jeon2020recent}. This regime dependence is formalized through the Lewis number in Section~2.3, and it provides the criterion by which the applicability of the CAFT model to rich hydrogen mixtures is assessed in this study.

\begin{table}[!t]
\centering
\caption{Summary of numerical studies on the flammability limit.
Updated and extended from Ref.~\cite{jeon2020recent}.}
\label{tab:numerical_studies}
\footnotesize
\resizebox{\textwidth}{!}{%
\begin{tabular}{clllc}
\toprule
Year & Researchers & Numerical method & Mixture & Limit \\
\midrule
1991 & Lakshmisha et al.~\cite{lakshmisha1991flammability}   & 1D/planar/non-adiabatic/unsteady            & $\mathrm{CH_4}$--air & LFL \\
1992 & Law et al.~\cite{law1992unified}                 & 1D/planar/non-adiabatic/steady              & $\mathrm{H_2}$--air, $\mathrm{CH_4}$--air & Both \\
1993 & Giovangigli et al.~\cite{giovangigli1993application} & 1D/planar/adiabatic/steady                & $\mathrm{H_2}$--air, $\mathrm{CH_4}$--air & Both \\
1998 & Ju et al.~\cite{ju1998flame}                   & 1D/stretched/non-adiabatic/steady              & $\mathrm{CH_4}$--air & LFL \\
2008 & Van den Schoor et al.~\cite{van2008calculation} & 1D/planar and spherical/non-adiabatic/steady and unsteady & $\mathrm{CH_4}$--air & UFL \\
2009 & Shoshin et al.~\cite{shoshin2009extinction}         & 2D/stretched/non-adiabatic/unsteady         & $\mathrm{CH_4}$--air & LFL \\
2009 & Fernandez-Galisteo et al.~\cite{fernandez2009one} & 1D/planar/adiabatic/steady/1-step      & $\mathrm{H_2}$--air & LFL \\
2015 & Hernandez-Perez et al.~\cite{hernandez2015formation} & 2D/stretched/non-adiabatic/steady         & $\mathrm{H_2}$--$\mathrm{CH_4}$--air & LFL \\
2018 & Zhou et al.~\cite{zhou2018experimental}           & 2D/stretched/non-adiabatic/steady           & $\mathrm{H_2}$--$\mathrm{CH_4}$--air & LFL \\
2018 & Zhou et al.~\cite{zhou2018effect}         & 2D/stretched/non-adiabatic/steady           & $\mathrm{H_2}$--air, $\mathrm{CH_4}$--air, $\mathrm{H_2}$--$\mathrm{CH_4}$--air & LFL \\
2018 & Yakovenko et al.~\cite{yakovenko2018large}     & 2D/stretched/non-adiabatic/unsteady         & $\mathrm{H_2}$--air & LFL \\
2019 & Bertolino et al.~\cite{bertolino2019prediction}     & 1D/planar/non-adiabatic/steady              & $\mathrm{CH_4}$-, $\mathrm{CH_3OH}$-, $\mathrm{C_2H_6}$-, $\mathrm{C_3H_8}$--air, etc. & Both \\
2021 & Jeon et al.~\cite{jeon2021identification}             & 2D/stretched/non-adiabatic/steady           & $\mathrm{H_2}$--air & LFL \\
2025 & Xie and Zhang~\cite{xie2025extinction}              & 1D/stretched/non-adiabatic/steady           & $\mathrm{NH_3}$--air, $\mathrm{NH_3}$--$\mathrm{H_2}$--air & Both \\
2025 & Yakovenko et al.~\cite{yakovenko2025downward}     & 2D/non-adiabatic/unsteady              & $\mathrm{H_2}$--air & LFL \\
\bottomrule
\end{tabular}%
}
\end{table}

\subsection{Lewis Number Effect on Flame Dynamics}

The Lewis number is defined as $Le=\alpha/D_{k,\mathrm{mix}}$, where $\alpha=\lambda/(\rho c_p)$ is the thermal diffusivity of the mixture and $D_{k,\mathrm{mix}}$ is the mixture-averaged mass diffusivity of the deficient reactant $k$. Both quantities must therefore be evaluated at the same temperature, pressure, and mixture composition. When $Le<1$, preferential mass diffusion of the deficient reactant relative to heat diffusion generally promotes diffusional--thermal instability, whereas $Le>1$ tends to suppress it. The associated stability trends have been observed experimentally in premixed $\mathrm{H_2/CO/air}$ mixtures and in expanding hydrogen/propane spherical flames \cite{burbano2011laminar,law2005cellular}. In the present analysis, the transport properties are evaluated using the mixture-averaged transport model in Cantera 3.2 at 1~atm \cite{goodwin2025cantera}. The \texttt{h2o2.yaml} mechanism is used for $\mathrm{H_2}$--air mixtures, whereas \texttt{gri30.yaml} is used for CO--air mixtures.

\begin{figure}[!t]
    \centerline{\includegraphics[width=\columnwidth]{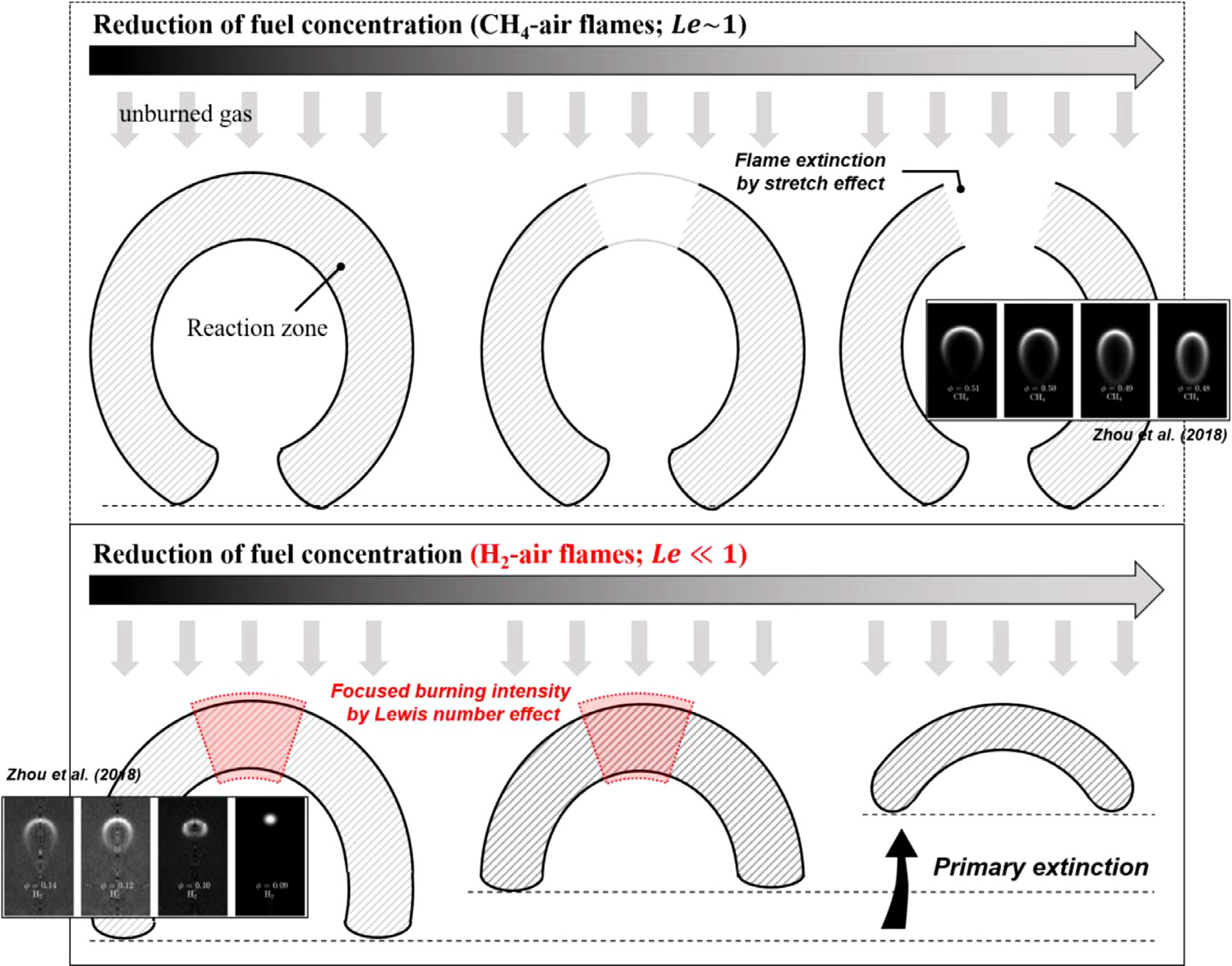}}
    \caption{Extinction mechanisms of lean $\mathrm{CH_4}$--air ($Le \approx 1$) and $\mathrm{H_2}$--air ($Le \ll 1$) flames: methane flames extinguish at the flame tip by the stretch effect, whereas hydrogen flames undergo primary extinction at the trailing edge. Adapted from Jeon et al.~\cite{jeon2021identification}.}
    \label{fig:extinction}
\end{figure}

In hydrogen--air mixtures, the Lewis number varies strongly with equivalence ratio. This behavior arises from the exceptionally high molecular diffusivity of $\mathrm{H_2}$ and the substantial increase in mixture thermal diffusivity as the hydrogen fraction increases. Under lean conditions ($\phi<1$), hydrogen is the deficient reactant. A near-limit lean mixture contains only a small fraction of hydrogen and therefore retains air-like bulk thermal properties. For 4.1~vol.\% $\mathrm{H_2}$ in air at 298~K, $\alpha_{\mathrm{mix}}=2.53\times10^{-5}~\mathrm{m^2\,s^{-1}}$, whereas $D_{\mathrm{H_2,mix}}=8.07\times10^{-5}~\mathrm{m^2\,s^{-1}}$. The resulting Lewis number is therefore $Le\approx0.31$. At the same mixture composition, this value changes only slightly when the initial temperature is increased to 383~K. For such low-$Le$ hydrogen flames, preferential diffusion of $\mathrm{H_2}$ can locally enrich the positively stretched leading region of the flame and enhance the local burning intensity \cite{zhou2018effect,jeon2021identification}. Jeon et al.\ demonstrated through stabilized-flame simulations that this preferential-diffusion effect strengthens burning near the flame tip. Consequently, the primary extinction of near-limit lean hydrogen flames occurs at the trailing edge through conductive heat loss (Fig.~\ref{fig:extinction}) \cite{jeon2021identification}. This behavior contrasts with the corresponding near-limit methane flames, for which extinction is initiated near the flame tip as local flame propagation is weakened by stretch.

Under fuel-rich conditions ($\phi>1$), the deficient reactant shifts from $\mathrm{H_2}$ to $\mathrm{O_2}$, and both transport properties must be re-evaluated for the hydrogen-rich mixture. For 75~vol.\% $\mathrm{H_2}$ in air at 298~K, the excess hydrogen increases the mixture thermal diffusivity to $\alpha_{\mathrm{mix}}=9.31\times10^{-5}~\mathrm{m^2\,s^{-1}}$. The hydrogen-rich background also increases the mixture-averaged oxygen diffusivity to $D_{\mathrm{O_2,mix}}=4.26\times10^{-5}~\mathrm{m^2\,s^{-1}}$, approximately twice its value in an air-like mixture. These increases partially compensate, yielding $Le\approx2.19$ at 298~K. At the same mixture composition, the corresponding value at 383~K is 2.16. The Lewis number is therefore only weakly dependent on temperature over this range. Within the dilute ideal-gas, mixture-averaged transport approximation, it is also approximately pressure-independent because both $\alpha$ and $D_{k,\mathrm{mix}}$ exhibit the same leading inverse-pressure dependence.

This trend is consistent with the Lewis-number variation reported by Burbano et al.\ \cite{burbano2011laminar}. Their hydrogen-flame curve reaches approximately $Le=1.9$ at $\phi=2$, whereas the present mixture-averaged evaluation gives $Le=2.04$ at the same equivalence ratio. The calculated Lewis number continues to increase beyond their reported range, reaching $Le=2.13$ at $\phi=3$ and $Le=2.19$ at the 75~vol.\% $\mathrm{H_2}$ composition, corresponding to $\phi\approx7.1$. Rich hydrogen--air flames are therefore substantially less susceptible to diffusional--thermal instability than their lean counterparts. This property is not limited to dry mixtures. Across the FITS boundary compositions of Section~3, with steam fractions of 5--50\%, the mixture-averaged Lewis number of oxygen remains between 1.35 and 2.10. For the rich-phase accident compositions of Section~4 it is 1.9--2.0.

The near-unity Lewis number of CO--air mixtures has a molecular transport origin. CO and $\mathrm{N_2}$ have nearly identical molecular masses and similar molecular transport characteristics. Replacing nitrogen with CO therefore produces only modest changes in both the mixture thermal diffusivity and the relevant reactant mass diffusivities over a wide range of compositions. On the lean side, CO is the deficient reactant and diffuses through an air-like background. On the rich side, $\mathrm{O_2}$ becomes the deficient reactant and diffuses through a CO-rich background whose transport characteristics remain similar to those of an $\mathrm{N_2}$-rich mixture. The mixture-averaged evaluation consequently gives $Le=1.06$--$1.07$ from the lean limit (12.5~vol.\% CO) to the rich limit (74~vol.\% CO). This near-constancy is consistent with Burbano et al., who reported $Le\approx1.1$ for CO--air flames over a wide range of equivalence ratios \cite{burbano2011laminar}. These fuel-specific Lewis-number characteristics provide an important background for the discussion of CAFT model applicability presented in the subsequent sections.

\subsection{CAFT Model}
 
The CAFT model, introduced by Egerton  and Zabetakis, is based on the empirical observation that the adiabatic flame temperature at the flammability limit is approximately constant \cite{egerton1953limits, zabetakis1965flammability}. Under the adiabatic assumption, all heat released by combustion is used solely to raise the temperature of the products. The energy balance then yields the adiabatic flame temperature. If a nearly constant threshold temperature can be identified for a class of fuels and mixture conditions, the flammability limit becomes predictable. The limit is obtained by inversely calculating the fuel concentration that reaches this critical temperature.
 
Numerous studies have confirmed that the CAFT remains relatively constant at the lean flammability limit of hydrocarbon fuels. Macek reported CAFT values in the range of approximately 1480--1650 K for hydrocarbons from methane to n-octane \cite{mavcek1979flammability}. Vidal et al.\ verified the consistency across chemical families, with average values of 1590 K for paraffinic hydrocarbons and 1600 K for unsaturated hydrocarbons \cite{vidal2006evaluation}. Mashuga and Crowl demonstrated that the flammable envelope of methane and ethylene could be reasonably predicted using a CAFT criterion of 1200 K \cite{mashuga1999flammability}. Wan et al.\ reported that the constancy of CAFT was maintained over initial temperatures of 300--1200 K and pressures of 1--100 bar \cite{wan2015theoretical}. The reasonable accuracy of the CAFT model for hydrocarbon fuels has been associated with Lewis numbers relatively close to unity, which limit preferential-diffusion effects \cite{wan2015theoretical}. It has also been linked to the CO-to-$\mathrm{CO_2}$ oxidation pathway, which becomes self-sustaining near 1400 K and provides a characteristic thermal threshold \cite{mashuga1999flammability, wan2015theoretical}.

A theoretical basis for relating CAFT to the actual flame temperature is provided by hydrogen flame-ball theory \cite{sanchez2014recent}. Consider a stationary spherical flame ball in which radiative heat loss is neglected and the transport properties are approximately constant. Integration of the energy and species conservation equations then relates the peak reaction-layer temperature $T_{\mathrm{peak}}$ to the calculated adiabatic flame temperature $T_{\mathrm{CAFT}}$ as follows.

\begin{equation}
\frac{T_{\mathrm{peak}}-T_u}{T_{\mathrm{CAFT}}-T_u} = \frac{\rho\, c_p^{*}\,\widetilde{D}_{\mathrm{H_2}}}{\lambda} = \frac{1}{\widetilde{Le}_{\mathrm{H_2}}}
\label{eq:caft_flameball}
\end{equation}
where $T_u$ is the unburned-gas temperature, $\rho$ is the mixture density, $c_p^{*}$ is the average specific heat of the mixture, $\widetilde{D}_{\mathrm{H_2}}$ is the effective hydrogen mass diffusivity, $\lambda$ is the thermal conductivity, and $\widetilde{Le}_{\mathrm{H_2}}$ is the effective Lewis number of hydrogen. For lean hydrogen--air mixtures near the flammability limit, $\widetilde{Le}_{\mathrm{H_2}}$ is approximately 0.3 \cite{sanchez2014recent, jeon2020recent}. This value is consistent with the mixture-averaged evaluation in Section~2.3. This proportionality was recently invoked to explain the rationale of the CAFT approach for hydrogen \cite{jeon2020recent}. Eq.~\ref{eq:caft_flameball} therefore shows that CAFT should not be interpreted as the actual peak reaction-zone temperature. Instead, the temperature rises $T_{\mathrm{peak}}-T_u$ and $T_{\mathrm{CAFT}}-T_u$ are proportional through the preferential-diffusion factor $1/\widetilde{Le}_{\mathrm{H_2}}$. If the effective Lewis number varies only weakly within a near-limit regime, a nearly constant critical peak flame temperature is therefore mapped onto a nearly constant CAFT. Flame balls are a structure of low-Lewis-number mixtures. In mixtures with $Le \geq 1$, near-limit flames instead remain quasi-planar, and the peak flame temperature remains bounded by the adiabatic value. In that regime the CAFT approximates the actual flame temperature directly. This distinction anticipates the rich-hydrogen conditions examined in this study.
 
This flame-ball interpretation also provides a connection between the thermal and kinetic criteria for hydrogen flammability. In reduced-kinetic flame-ball analyses, sustained combustion ceases when the reaction-layer peak temperature falls toward the kinetic crossover temperature \cite{fernandez2012flammability}. Below this temperature, radical production cannot maintain the reaction. Thus, the empirical constancy of CAFT can be interpreted as an indirect representation of a more fundamental threshold in the local reaction-zone temperature. Recently, a first-principles analysis of hydrogen--air concentration limits provided a complementary theoretical basis for this empirical constancy \cite{plaksin2025hydrogen}. The near-limit flame temperature is governed by the kinetic crossover temperature, at which the chain-branching and chain-termination rates balance. This crossover temperature depends only weakly on the mixture stoichiometry.

The CAFT at the lean flammability limit of hydrogen mixtures is approximately 600 K, which is notably lower than those of hydrocarbon fuels \cite{jeon2021mechanistic,jeon2019flammability}. Importantly, this relatively low CAFT does not imply that the local reaction-zone temperature of a lean hydrogen flame is only approximately 600 K. According to Eq.~\ref{eq:caft_flameball}, the low effective Lewis number of hydrogen amplifies the local peak-temperature rise relative to the corresponding adiabatic temperature rise. Jeon et al.\ calculated the CAFT for various lean-limit hydrogen mixtures and classified them into two groups based on the CNAFT coefficient ($\pi=\alpha/C$) \cite{jeon2019flammability}. For Group 1 mixtures ($\mathrm{H_2}$-air, $\mathrm{H_2}$-air-$\mathrm{N_2}$, $\mathrm{H_2}$-air-Ar, $\mathrm{H_2}$-air-$\mathrm{CO_2}$), in which the coefficient was below the reference value, the CAFT remained relatively constant at approximately 600--620 K with prediction errors within 6\%. However, for Group 2 mixtures (initial temperatures above 100~$^\circ$C, helium dilution, or steam dilution), in which the coefficient exceeded the reference value, the CAFT increased up to 913 K, with prediction errors reaching 40\% \cite{jeon2019flammability}. The accuracy degradation was particularly pronounced for $\mathrm{H_2}$-air-steam mixtures.
 
This behavior originates from the adiabatic assumption inherent to the CAFT model. In real flames, a fraction of the heat generated in the reaction zone is lost through radiation and conduction, and the magnitude of this heat loss depends on the thermal and radiative properties of the mixture \cite{jeon2021mechanistic,jeon2019flammability}. When radiating species such as steam are present, the volumetric radiation rate increases, amplifying the heat loss and causing the CAFT prediction to deviate from the actual flame behavior \cite{jeon2021mechanistic}. In terms of Eq.~\ref{eq:caft_flameball}, such non-adiabatic effects violate the assumptions behind the proportional relationship between $T_{\mathrm{peak}}$ and $T_{\mathrm{CAFT}}$. The direct applicability of a constant-CAFT criterion is then limited.
 
Jeon et al.\ developed the CNAFT model by relaxing the adiabatic assumption of the CAFT model, incorporating an indirect radiation (conduction) heat loss term into the energy balance equation \cite{jeon2021mechanistic,jeon2019flammability}. The heat loss is estimated through a linear relationship with the CNAFT coefficient, and for steam-containing mixtures, the radiative effect of steam is additionally accounted for via the optically thin approximation \cite{jeon2021mechanistic}. This model demonstrated a maximum relative error of approximately 13\% when compared against LFL experimental data for a wide range of mixtures including $\mathrm{H_2}$--air, $\mathrm{H_2}$--air--He, $\mathrm{H_2}$--air--$\mathrm{N_2}$, $\mathrm{H_2}$--air--Ar, $\mathrm{H_2}$--air--$\mathrm{CO_2}$, and $\mathrm{H_2}$--air--steam, representing a substantial improvement over the 40\% maximum error of the original CAFT model \cite{jeon2021mechanistic}.
 
The physical basis of the CNAFT model is supported by mechanistic analysis of lean hydrogen flame extinction. Jeon et al.\ confirmed through numerical simulations that conductive heat loss at the trailing edge exceeds the direct radiative heat loss by more than a factor of 20, corroborating the validity of the CNAFT model in considering only the indirect radiation mechanism \cite{jeon2021identification}.
 
For the lean hydrogen mixtures targeted by the CAFT and CNAFT approaches, model accuracy is strongly influenced by the heat loss during flame propagation. It is influenced equally by the mixture transport properties that determine preferential diffusion and the local flame structure. The reduced accuracy reported for lean hydrogen mixtures with radiating diluents reflects the specific role of indirect radiation at the trailing edge of lean-limit flames, where preferential diffusion concentrates the burning intensity at the flame tip \cite{jeon2021identification}. Whether such heat-loss-driven deviations also dominate the rich-hydrogen UFL regime remains an open question that requires separate investigation. That regime differs fundamentally from the lean side in stoichiometry, in the dominant three-body termination kinetics, and in Lewis number effects.

\subsection{Research Gap}

Although the CAFT framework has been studied extensively for flammability-limit prediction, its application to hydrogen has remained largely focused on lean conditions. CAFT-based UFL predictions for hydrogen have been reported mainly for dry $\mathrm{H_2}$--air mixtures \cite{terpstra2012flammability,liaw2016model}. The CNAFT model, developed to account for heat-loss effects, has been applied primarily to LFL prediction under lean conditions \cite{jeon2019flammability,jeon2021mechanistic}. For hydrocarbon fuels, previous studies have shown that CAFT can also be useful for UFL prediction \cite{van2008calculation,shu2015flame}. Fuel-rich conditions, however, introduce additional uncertainties associated with incomplete combustion and product composition \cite{shu2015flame}. These findings do not establish whether the same thermal criterion remains applicable to rich hydrogen flames.

The key unresolved issue is therefore twofold. It concerns whether the thermal criterion underlying the CAFT model remains physically appropriate under rich-hydrogen conditions. It also concerns whether a constant threshold temperature can classify flammability consistently with measured hydrogen UFL data. As discussed in Section~2.3, the deficient reactant changes from $\mathrm{H_2}$ to $\mathrm{O_2}$ as the mixture becomes fuel-rich. The Lewis number correspondingly increases from $Le<1$ on the lean side to $Le>1$ under sufficiently rich conditions. The resulting transport behavior differs fundamentally from that of lean hydrogen flames, where strong preferential diffusion governs the local burning structure and extinction. Rich hydrogen flames also involve different chemical kinetics, including the increasing importance of three-body termination reactions. Despite these changes in transport and chemistry, the applicability of the CAFT concept to the rich-hydrogen regime has not been examined mechanistically.

Experimental validation under conditions relevant to nuclear severe accidents also remains limited. The FITS experiments provide the representative UFL database for $\mathrm{H_2}$--air--steam mixtures \cite{marshall1986hydrogen}. These measurements have been widely used as the basis for hydrogen flammability criteria in severe-accident analyses \cite{jeon2020recent}. To the best of the authors' knowledge, no study has combined a mechanistic assessment of rich-side applicability with a quantitative validation against the FITS database. Consequently, both the physical basis and the practical classification capability of the CAFT model in this regime remain insufficiently established.

The present study addresses these gaps through two complementary approaches. First, the physical applicability of the CAFT model to rich hydrogen flames is examined in terms of transport characteristics, heat-loss behavior, and the kinetic basis of the critical temperature. Particular attention is given to the Lewis-number transition between lean and rich mixtures and its implications for the underlying thermal criterion. Second, the CAFT is evaluated along the measured FITS UFL boundary to test the constancy of the limit temperature. The minimum limit value then defines a conservative flammability threshold, and its classification performance and applicable envelope are quantified. The validated criterion is subsequently applied to evaluate hydrogen combustion potential during a NuScale SMR severe accident scenario.

\section{CAFT Methodology and Validation under Hydrogen-rich Conditions}

The CAFT methodology is grounded in fundamental thermodynamic principles. It is therefore applicable to arbitrary gas compositions. In severe accident environments, multi-component mixtures of hydrogen, oxygen, steam, and nitrogen evolve continuously as the accident progresses. Empirical flammability data tied to fixed reference conditions may then be insufficient for a comprehensive assessment \cite{marshall1986hydrogen, plys1993hydrogen}. Thermodynamic criteria such as CAFT can instead evaluate flammability over a wide range of mixture states.

The fuels for which the CAFT model performs well share a common transport characteristic. Their Lewis numbers are close to or above unity. For hydrocarbon fuels such as methane and propane, the Lewis number remains near unity. The same holds for CO at every equivalence ratio, as shown in Section~2.3 \cite{burbano2011laminar}. For these fuels the limit CAFT stays within approximately 1480--1650 K \cite{mavcek1979flammability}. In contrast, the accuracy deteriorates markedly for lean hydrogen mixtures with $Le \ll 1$ \cite{jeon2019flammability, jeon2021mechanistic}.

Rich hydrogen mixtures fall on the favorable side of this classification. As discussed in Section~2.3, the deficient reactant shifts to $\mathrm{O_2}$ and the Lewis number exceeds unity \cite{burbano2011laminar}. Rich hydrogen flames therefore belong to the Lewis-number regime in which the CAFT model is known to perform well. This correspondence motivates the application of the CAFT model under rich-hydrogen conditions.

The Lewis number alone, however, does not determine the accuracy of the CAFT model. Heat loss associated with the adiabatic assumption exists independently of the Lewis number. For steam-containing mixtures, radiative heat loss from this participating species additionally comes into play \cite{jeon2021mechanistic}. Whether a constant threshold temperature actually holds along the measured rich-hydrogen boundary therefore requires an empirical test. This section provides that test using the FITS experimental database. The constancy of the limit temperature is examined, and its minimum value is adopted as a conservative flammability threshold.

\subsection{CAFT Methodology}

The CAFT methodology is based on the empirical observation that the adiabatic flame temperature near the flammability limits of a given fuel-oxidizer mixture assumes a nearly constant value \cite{egerton1953limits, zabetakis1965flammability}. The adiabatic flame temperature is calculated from the energy balance relation given in Eq.~\ref{eq:caft}:

\begin{equation}
\label{eq:caft}
\begin{aligned}
&\sum_{reactants}n_i\left [\Delta H_{f,i}^{0}+\bar{c}_{p,i}(T_{i}-T_{ref}) \right ]- \\
&\sum_{products}n_i\left [\Delta H_{f,i}^{0}+\bar{c}_{p,i}(T_{CAFT}-T_{ref}) \right ] = 0
\end{aligned}
\end{equation}

Under adiabatic conditions, all heat released by the combustion reaction is entirely consumed in raising the temperature of the products. In this process, the standard enthalpy of formation of each chemical species ($\Delta H^0_{f,i}$), the reference temperature ($T_{ref}$, typically 298 K), and the mean specific heat of the mixture ($\bar{c}_{p,i}$) serve as the key thermodynamic variables.

According to Arrhenius theory, the peak flame temperature observed at the flame front is the governing factor that determines the overall combustion reaction rate. If the heat generated by chemical reactions cannot compensate for heat losses to the surroundings, the flame cannot propagate continuously and ultimately extinguishes. Therefore, a minimum amount of combustion heat must be generated for sustained flame propagation, and a critical temperature exists that satisfies this requirement. The CAFT methodology evaluates flammability by calculating the adiabatic flame temperature for a given gas mixture and determining whether this value exceeds the critical temperature required for flame propagation \cite{mavcek1979flammability, mashuga1999flammability}.

Independent literature values support the existence of a quasi-constant limit temperature on the rich side. Terpstra derived 1168 K as the adiabatic flame temperature at the measured rich limit of hydrogen in air (74.1 vol\% for upward propagation at 20~$^\circ$C) \cite{terpstra2012flammability}. Assuming this value constant reproduces the measured rich limits within approximately 1\% up to 150~$^\circ$C, a range that encompasses the FITS initial temperature of 383 K. The corresponding constant-value assumption on the lean side deviates by up to 29\% over the same range \cite{terpstra2012flammability}. This asymmetry independently corroborates the Lewis-number-based expectation of Section~2.3 that the limit flame temperature is quasi-constant on the rich side. Kumar likewise reported a rich-limit flame temperature of approximately 1140 K for hydrogen--oxygen--diluent mixtures \cite{kumar1985flammability}. In the present study, the operational threshold is not taken from these literature values. It is instead derived from the FITS boundary data themselves, as described in Section~3.3. The literature values serve as an independent check on the data-derived threshold.

\subsection{Calculation Methodology}

\begin{figure}[!t] \centerline{\includegraphics[width=\textwidth]{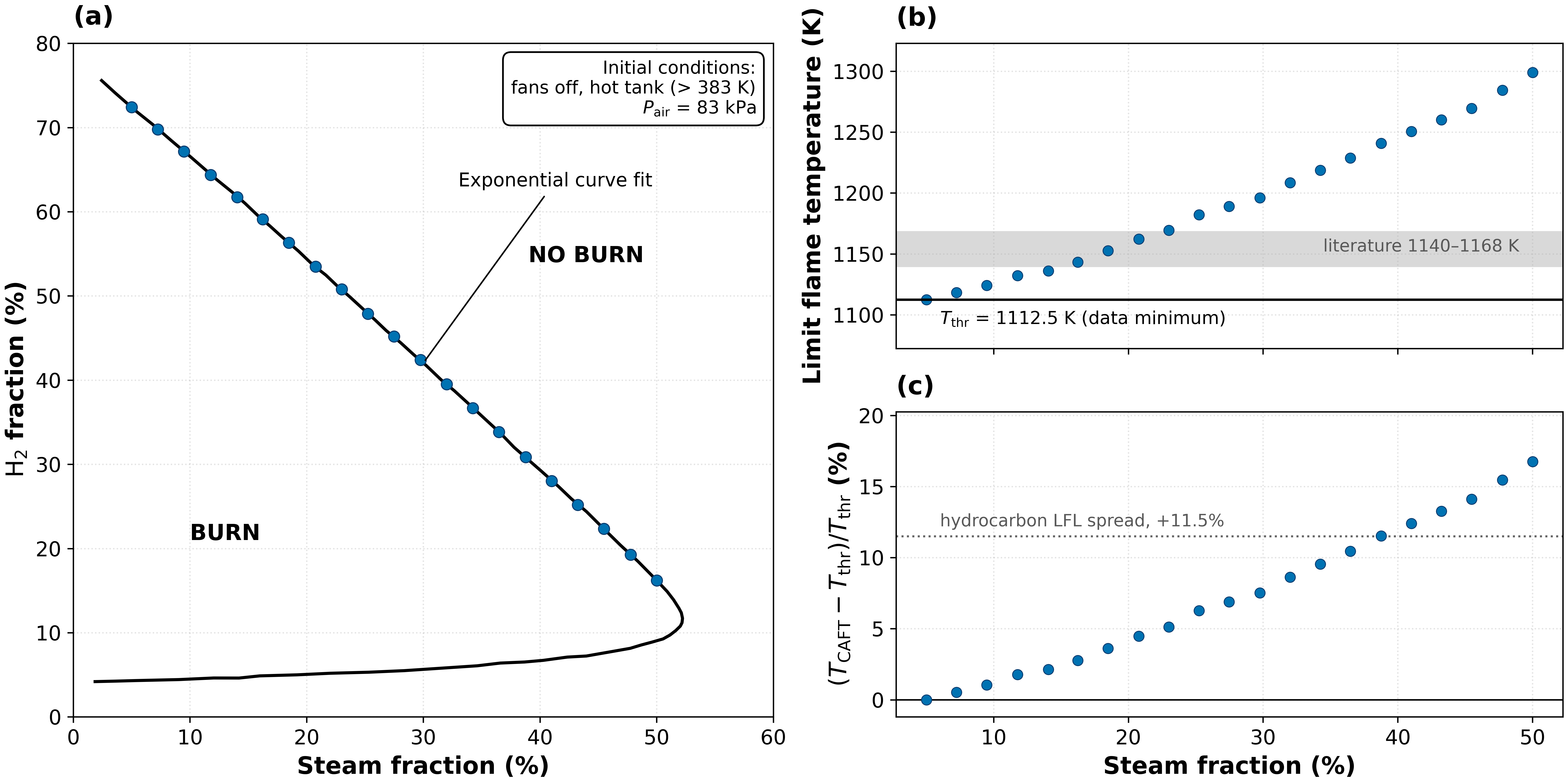}} \caption{Validation of the CAFT-based flammability criterion against the FITS rich-limit data. (a) FITS flammability boundary replotted from Ref.~\cite{marshall1986hydrogen}, with the 21 sampled points on the rich branch. (b) Calculated limit flame temperatures along the boundary. The solid line marks the data-derived threshold of 1112.5 K. The shaded band shows independently reported critical temperatures of 1140–1168 K \cite{kumar1985flammability, terpstra2012flammability}. (c) Relative deviation of the limit temperature from the threshold. All deviations are non-negative because the threshold is defined as the minimum of the dataset, so misclassification is confined to the conservative side. The dotted line indicates the 11.5\% relative spread of hydrocarbon LFL flame temperatures, derived as $(1650-1480)/1480$ from the 1480--1650~K range of Ma\v{c}ek~ \cite{mavcek1979flammability}.} \label{fig:fits_validation} \end{figure}

To test the constancy of the limit flame temperature under rich-hydrogen conditions and to derive a flammability threshold, the FITS experimental database from the U.S. NRC was employed \cite{marshall1986hydrogen}. The FITS experiments were conducted to investigate the combustion characteristics of $\mathrm{H_2}$–air–steam mixtures that may form inside the containment building during a nuclear severe accident. A total of 239 combustion tests systematically examined the effects of steam concentration, mixture ratio, initial pressure, and temperature on combustion pressure rise and flame behavior. In particular, the FITS experiments include conditions with high hydrogen concentrations, making them suitable for validating flammability assessment methodologies in rich-hydrogen environments.
In this study, 21 representative data points along the measured UFL boundary were sampled from the FITS experiments and applied to the CAFT model (Fig.~\ref{fig:fits_validation}(a)). This enabled the calculation of the limit flame temperature for each boundary composition. The sampled data encompass steam mole fractions ranging from approximately 5\% to 50\%, with corresponding equivalence ratios spanning from approximately 1.14 to 7.64. All 21 data points correspond to hydrogen-rich conditions with equivalence ratios above unity. The initial temperature and air partial pressure were set to the experimental baseline values of 383 K and 83 kPa, respectively.

The equivalence ratio for each FITS experimental data point is defined as follows:

\begin{equation}
\phi =\frac{\left ( X_{H_2}/X_{O_2} \right )_{actual}}{\left ( X_{H_2}/X_{O_2} \right )_{stoichiometric}}
\end{equation}

From the stoichiometric reaction $2H_2+O_2 \to 2H_2O$, the stoichiometric fuel-to-oxidizer ratio is $(H_2/O_2)_{stoichiometric}=2$. Nitrogen is treated as an inert diluent and is therefore excluded from the oxidizer in the equivalence ratio calculation. Similarly, steam is treated as a non-participating diluent.

The adiabatic flame temperature was calculated using Eq.~\ref{eq:caft}. Under rich-hydrogen conditions near the UFL, oxygen is the deficient reactant. Accordingly, an oxygen-limited reaction model was applied to determine the product composition. In this model, the hydrogen combustion reaction ($H_2+1/2O_2\to H_2O$) proceeds until all available oxygen is consumed, and unreacted hydrogen remains in the products. The adiabatic flame temperature was then determined by establishing the energy balance under adiabatic conditions using the reaction enthalpy and the specific heat of each gaseous species. Full-equilibrium adiabatic flame temperature calculations differ from this product model by less than 0.1\% over the range of interest. Dissociation effects are therefore negligible for these hydrogen-rich products.

The CAFT was evaluated for each of the 21 boundary compositions using Eq.~\ref{eq:caft} with the oxygen-limited reaction model. The threshold temperature was then defined as the minimum of these limit values. This minimum, 1112.5 K, occurs at the lowest sampled steam fraction. The criterion applies to the fuel-rich branch ($\phi > 1$), where hydrogen addition beyond stoichiometry monotonically lowers the CAFT. Any flammable state at a given steam fraction therefore has a CAFT no lower than its local boundary value. Since the threshold is the minimum over all boundary values, the condition $T_{\mathrm{CAFT}} \geq T_{\mathrm{Threshold}}$ captures every flammable state represented by the FITS boundary. Misclassification is confined to the conservative side, where some non-flammable states just above the boundary are still flagged.

Criterion consistency is assessed in temperature space rather than by back-calculating the UFL, for the following reasons. First, the physical claim of the CAFT model concerns a thermal state. Flame propagation ceases when the flame temperature falls below a critical value, so the constancy of the limit temperature is the direct test. Previous evaluations of the CAFT and CNAFT models adopted the same measure \cite{jeon2019flammability, jeon2021mechanistic}. Second, back-calculating a UFL requires a composition path along which hydrogen is varied, such as fixing the relative proportions of the background gases. This path is a computational convention rather than an experimental process. Composition-space errors therefore mix model error with the choice of convention. Third, the inversion becomes highly sensitive to temperature errors near the rich limit. The effective temperature gradient along the composition path decreases from approximately 26 K per percentage point of hydrogen at 5\% steam to 9 K at 50\% steam. A modest temperature deviation is thus amplified into a large composition deviation at high steam fractions. Finally, the application in Section~4 uses the model as a temperature-threshold classifier on known mixture states. Validation is therefore performed in the same space as the application.

\subsection{Limit Temperature Constancy and Threshold Derivation}

Fig.~\ref{fig:fits_validation}(b) shows the calculated limit flame temperatures for the 21 boundary compositions. The values range from 1112.5 K to 1298.8 K over steam fractions of 5--50\%. The variation stays within 17\% of the minimum, with a coefficient of variation of approximately 5\%. This quasi-constancy is comparable to the hydrocarbon LFL band of 1480--1650 K, which corresponds to a relative spread of approximately 11.5\% \cite{mavcek1979flammability}. It contrasts sharply with the lean-hydrogen side, where the limit CAFT rises from approximately 600 K to 913 K \cite{jeon2019flammability}. The minimum occurs at the lowest sampled steam fraction and defines the threshold $T_{\mathrm{Threshold}} = 1112.5$ K. Every boundary composition therefore satisfies $T_{\mathrm{CAFT}} \geq T_{\mathrm{Threshold}}$ by construction. This property is a designed feature rather than an independent test. The empirical test lies in the quasi-constancy of the limit temperature, which fails on the lean side.

Fig.~\ref{fig:fits_validation}(c) shows the relative deviation of the limit temperature from the threshold. The deviation grows systematically with steam fraction rather than scattering randomly. This trend is consistent with the radiative heat loss of steam discussed in Section~2.4 \cite{jeon2021mechanistic}. The maximum deviation of +16.7\% corresponds to 186 K and stays within the 11.5\% hydrocarbon-level spread for steam fractions below approximately 38\%. The deviation direction is uniformly conservative for flammability classification. The threshold also agrees with independently reported critical temperatures near the hydrogen UFL, approximately 1140 K by Kumar \cite{kumar1985flammability} and 1168 K by Terpstra \cite{terpstra2012flammability}. Both values lie within 5\% of the data-derived threshold. This agreement indicates that the threshold reflects a physical limit temperature rather than a fitting artifact.

The threshold derived here is applied in Section~4 to the time-resolved flammability assessment of the NuScale severe accident scenario. The validated envelope covers steam fractions of 5--50\%, hydrogen fractions of 16--72\%, an initial temperature of 383 K, and total pressures of approximately 2.5--3.7 bar. The boundary points themselves span this pressure range, and the limit temperature remains quasi-constant across it. The threshold also agrees within 5\% with literature values obtained near 1 bar. Both observations indicate a weak pressure dependence of the critical limit temperature.

\section{Application to NuScale SMR Severe Accident}
\subsection{NuScale SMR Design and Containment Characteristics}

The NuScale SMR represents one of the leading SMR designs, featuring enhanced economic efficiency through modular construction and strengthened passive safety functions. While conventional large PWR containment buildings are designed with large free volumes (approximately $5-8 \times 10^4~ m^3$) that are favorable for hydrogen dilution \cite{camp1983light}, the NuScale SMR adopts a vacuum containment vessel with a significantly smaller internal volume \cite{weber2020severe}. This design characteristic carries important implications for hydrogen combustion risk assessment in two respects.
First, the initial oxygen concentration within the containment vessel is very low. The vacuum containment design maintains the containment interior at sub-atmospheric pressure during normal operation, resulting in a very low oxygen concentration. Consequently, it has been widely recognized that even if hydrogen is generated, the conditions inside the containment are unfavorable for combustion \cite{weber2020severe}.
Second, the limited containment volume can lead to elevated hydrogen concentrations even for the same amount of hydrogen generation. During a severe accident, if large quantities of hydrogen are produced through zirconium oxidation reactions, hydrogen accumulates rapidly within the confined containment volume, potentially establishing rich-hydrogen conditions \cite{weber2020severe, a2019_assessment}.

However, recent studies have indicated that the NuScale SMR is not entirely free from hydrogen combustion concerns \cite{a2019_assessment, a2016_designspecific}. Severe accident simulation results show that the oxygen concentration within the containment can reach non-negligible levels, differing significantly from the original design expectations. When core cladding oxidation is less than 100\%, the relatively lower hydrogen production may actually create conditions favorable for flammability \cite{a2019_assessment}. Furthermore, the oxygen generated solely by radiolysis of coolant water has been shown to potentially satisfy the flammability conditions near the UFL \cite{a2016_designspecific,  a2025_safety}. In addition, the limited internal space of SMR containment vessels may constrain the installation of combustible gas control equipment such as PARs. Therefore, a rigorous flammability assessment specific to the rich-hydrogen conditions unique to SMRs is required. These conditions are distinct from the lean-hydrogen conditions of conventional large-scale NPPs.

\subsection{Severe Accident Scenario (LCC-05T-03)}

The severe accident scenario applied in this study is the LCC-05T-03 scenario presented in the ERI MELCOR analysis report for the NuScale SMR \cite{a2019_assessment}.

The LCC-05T-03 scenario assumes a rupture of the CVCS (Chemical and Volume Control System) piping inside the containment vessel of the NuScale module. Immediately following the rupture, reactor coolant is directly released into the containment vessel, causing an increase in containment pressure and temperature, and core uncovery due to coolant loss. Subsequently, zirconium oxidation of the fuel cladding proceeds, generating hydrogen. In the ERI simulation, hydrogen generation by steam-oxidation of the core structures begins at approximately 43 hours after accident initiation and continues until 49.38 hours, producing hydrogen in large quantities \cite{a2019_assessment}. Owing to the relatively small containment volume of the NuScale SMR, hydrogen accumulation rapidly depletes the oxygen concentration, causing the containment atmosphere to reach a virtually inert state during the early phase of the accident.

The ERI report performed a variant analysis of the LCC-05T-03 scenario to assess the potential for flammable conditions to develop. In this variant analysis, the ECCS valves (1 reactor vent valve and 1 reactor recirculation valve) were actuated at the 49.38-hour mark to halt further cladding oxidation. This measure was intended to terminate oxygen dilution caused by ongoing hydrogen generation and to establish an environment in which oxygen produced by radiolysis of the coolant could gradually accumulate. This allowed an assessment of whether the oxygen concentration could reach a level suitable for combustion, approximately 0.05 mole fraction. The ERI analysis selected the ECCS actuation time of 49.38 hours by a bounding back-calculation. The cumulative oxidation hydrogen at that time corresponds to approximately 19\% of the 93.3~kg hydrogen mass for 100\% oxidation of the active fuel cladding. Together with subsequent radiolytic oxygen generation, this choice drives the containment oxygen mole fraction to approximately 0.05 at the 72-hour mark~\cite{a2019_assessment}. It should be noted that hydrogen combustion events were intentionally suppressed in the ERI simulation until 72 hours. The flammability of the containment atmosphere prior to 72 hours was therefore outside the scope of the ERI assessment.

The key characteristics of the LCC-05T-03 scenario are as follows. First, the containment vessel is modeled as a closed single volume with no gas exchange with the external environment. Accordingly, the behavior of hydrogen, oxygen, nitrogen, and steam occurs entirely within the containment as a closed system. Second, oxygen is continuously generated over the long term through radiolysis. Even after the cessation of cladding oxidation, the oxygen concentration gradually increases with time.

The flammability assessment in the present study was limited to the first 72 hours following accident initiation. This is because hydrogen generation, oxygen accumulation, and the formation of flammable conditions predominantly occur during the early phase of the accident, and the possibility of implementing additional measures exists beyond this time frame.

\subsection{Gas Composition Evolution during Accident Progression}

\begin{figure}[!t]
    \centerline{\includegraphics[width=\columnwidth]{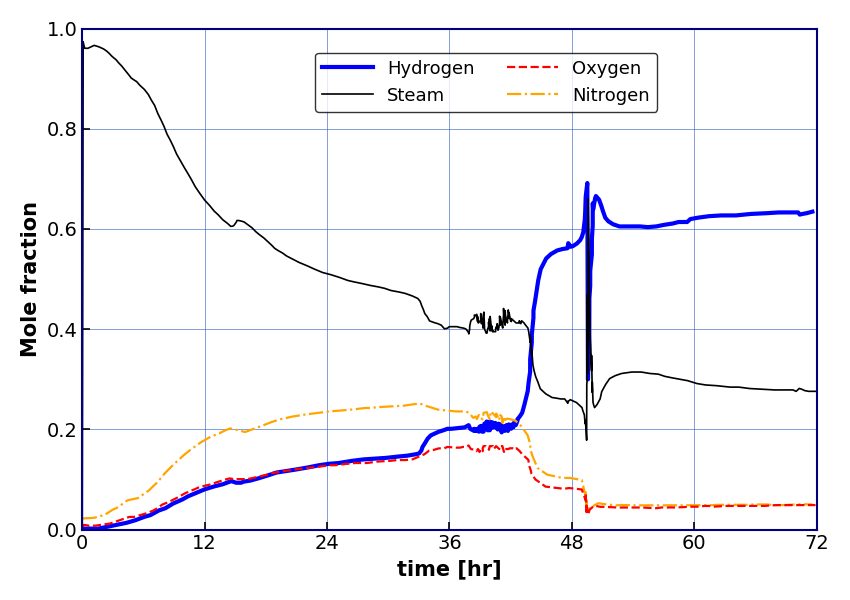}}
    \caption{Evolution of mole fractions for major gas species in the containment during the 
first 72 hours of the LCC-05T-03 variant scenario. Redrawn from the MELCOR results reported in Ref.~\cite{a2019_assessment}.}
    \label{fig:gas_evolution} 
\end{figure}

Fig.~\ref{fig:gas_evolution} shows the temporal evolution of mole fractions for the major gas species inside the containment vessel during the first 72 hours of the LCC-05T-03 variant scenario \cite{a2019_assessment}. 

During the early phase of the accident (0 to approximately 36 hours), the containment atmosphere is primarily composed of steam and nitrogen. The rising nitrogen and oxygen fractions during this phase reflect the initial air inventory and radiolytic oxygen generation. Nitrogen originates entirely from the initial air inventory because no accident process generates it. As discharged steam condenses inside the pool-cooled vessel, the mole fraction of this fixed air inventory increases. Radiolysis simultaneously adds oxygen and the stoichiometrically paired hydrogen~\cite{a2019_assessment}. The relatively small break area of the LCC-05T-03 scenario limits the steam dilution of the initial oxygen inventory, which is why this scenario yields the highest containment oxygen concentration among the scenarios considered in the ERI report \cite{a2019_assessment}. From approximately 43 hours onward, the zirconium oxidation reaction becomes vigorous, causing a sharp increase in the hydrogen mole fraction and a concurrent decrease in oxygen concentration. During the 44–48 hour interval, the hydrogen fraction rises sharply to approximately 45–57\%, while the oxygen fraction remains at approximately 8–10\%. In this interval, the steam fraction is approximately 25–31\%, which falls within the validated envelope of the criterion (steam fractions of 5--50\%, Section 3.3). This interval is the period in which a high oxygen fraction coexists with a hydrogen fraction that has not yet exceeded the upper flammability limit.

At 49.38 hours, the ECCS valves are opened, resulting in core re-flooding and the cessation of further hydrogen generation \cite{a2019_assessment}. The accompanying blowdown discharges the steam and hydrogen retained in the reactor pressure vessel into the containment, diluting the oxygen mole fraction from approximately 8\% to below 5\%. As the discharged steam subsequently condenses inside the containment vessel, the non-condensable hydrogen is enriched to approximately 63--66\%, while the steam fraction settles at approximately 24--31\%. Thereafter, oxygen generated by radiolysis of the coolant gradually accumulates, reaching an oxygen mole fraction of approximately 5\% (0.05) at the 72-hour mark. The ERI report considers this time point as the worst-case condition for flammability analysis.

\begin{figure}[!t]
    \centerline{\includegraphics[width=\columnwidth]{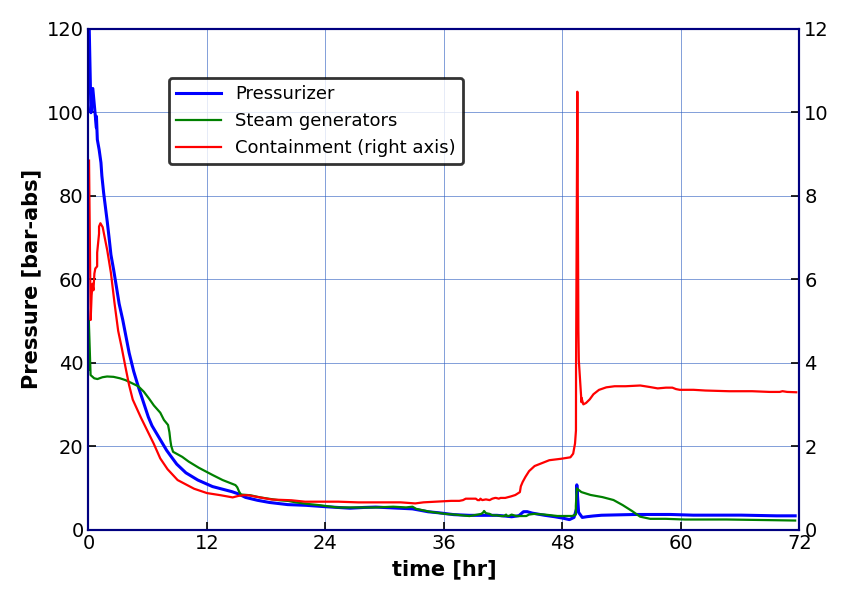}}
    \caption{Evolution of pressure in the containment 
during the first 72 hours of the LCC-05T-03 variant scenario. Redrawn from the MELCOR results
reported in Ref.~\cite{a2019_assessment}.}
    \label{fig:pressure} 
\end{figure}

Fig.~\ref{fig:pressure} presents the pressure histories of the reactor pressure vessel (RPV), steam generator (SG), and containment vessel (CNV) under the same scenario. At 49.38 hours, the opening of the ECCS valves establishes pressure equalization between the RPV and the containment, during which a transient increase in containment pressure is observed. Subsequently, cooling proceeds without additional hydrogen generation, and the system reaches a quasi-steady state, maintaining a relatively constant pressure.

According to the ERI report, the gas temperature inside the containment vessel in the LCC-05T-03 scenario was maintained at approximately 96.5~$^\circ$C \cite{a2019_assessment}. This temperature condition was applied in the flammability assessment conducted in the present study. The compositions in Table~\ref{tab:flammability_results} were digitized from the ERI report figures, and their sums deviate from 100\% by less than 0.5\%. The calculated flame temperature depends only on the mole ratios, so this deviation does not affect the results.

\subsection{Time-Dependent Flammability Assessment Results}

Since the hydrogen fraction increases sharply from approximately 44 hours after accident initiation, establishing rich-hydrogen conditions (equivalence ratio $\phi > 1$, with oxygen as the deficient reactant), the adiabatic flame temperature was calculated using gas composition data for each time interval after 44 hours, where the CAFT framework validated in Section 3 is applicable. It should be emphasized that the preceding hydrogen-lean phase is not thereby assessed as non-flammable. The steam fraction falls below the empirical inerting level of approximately 55\% at 22--24 hours. The mixture then passes through the stoichiometric composition at approximately 43 hours. Between these times the atmosphere contains approximately 13--25\% hydrogen, 13--17\% oxygen, and 40--50\% steam. These values satisfy conventional composition-based flammability criteria (oxygen at least approximately 5\%, steam below approximately 55\%, and hydrogen well above the lean flammability limit)~\cite{camp1983light, marshall1986hydrogen}. A quantitative assessment of this lean regime requires a lean-side non-adiabatic model such as CNAFT and lies outside the validated scope of the present rich-side framework. This extension is identified as future work (see Section 5). The adiabatic flame temperature, equivalence ratio, and flammability assessment results for each time interval are summarized in Table~\ref{tab:flammability_results}.

During the 44–48 hour interval, the calculated adiabatic flame temperatures exceed approximately 1500 K, substantially surpassing the threshold of 1112.5 K derived in Section~3.3. At the 44-hour mark, the adiabatic flame temperature reaches approximately 1835 K under conditions of approximately 10.4\% oxygen, 45\% hydrogen, and 31\% steam fractions. At the 46-hour and 48-hour marks, the temperatures remain well above the threshold at approximately 1553 K and 1536 K, respectively. Accordingly, this interval is assessed as a flammable region from a thermodynamic perspective. The steam fraction during this interval (approximately 25--31\%) falls well inside the validated envelope of the criterion (5--50\% steam). The containment total pressure in this interval (approximately 1.1--1.7 bar) lies below the FITS total pressures of 2.5--3.7 bar. However, the constant-pressure CAFT depends only on the mixture composition and the initial temperature under the ideal-gas assumption. The pressure difference therefore does not affect the threshold comparison. Moreover, the calculated flame temperatures exceed the threshold by more than 400 K. The conclusion of flammability is therefore robust even when accounting for the quantitative uncertainties of the CAFT model. This assessment is further supported by the screening criterion employed in the ERI report for long-term detonability evaluations (hydrogen fraction of at least 15\%, oxygen fraction of at least 5\%, and steam fraction of at most 35\%) \cite{a2019_assessment}: the containment atmosphere during the 44–48 hour interval satisfies all three conditions.

In the subsequent 50–72 hour interval, following the opening of the ECCS valves at 49.38 hours, the adiabatic flame temperature decreases sharply due to the dilution of oxygen by the blowdown and the accompanying enrichment of hydrogen. At the 50-hour mark, the adiabatic flame temperature drops to approximately 1006 K, falling below the threshold of 1112.5 K, and remains in the range of approximately 940–1033 K through 72 hours. Accordingly, the thermodynamic flammability criterion is not satisfied during this interval. Although oxygen gradually accumulates through radiolysis, reaching an oxygen mole fraction of approximately 5\% at 72 hours, the mixture remains deeply over-rich, with an equivalence ratio of approximately 6.8. The heat released by the limited amount of available oxygen is insufficient to raise the temperature of the hydrogen-dominated mixture above the threshold, and the atmosphere remains in an over-rich, non-flammable state. An oxygen-only reading would treat the approach to the 0.05 threshold as an emerging combustion concern beyond 72 hours. The thermodynamic margin instead stays negative because the mixture is deeply over-rich. The adiabatic isochoric complete combustion (AICC) evaluation in the ERI report supports the same conclusion at the 72-hour state \cite{a2019_assessment}. Because the threshold is anchored at the data floor, the criterion errs toward flagging flammability. A non-flammable verdict under this conservative criterion is therefore robust. The flammability verdicts in Table~\ref{tab:flammability_results} are moreover unchanged for any threshold between the data-derived 1112.5 K and the literature value of 1168 K.

In summary, the CAFT-based flammability assessment for the LCC-05T-03 scenario of the NuScale SMR identifies two distinctly different regimes over the course of the accident progression following the onset of rich-hydrogen conditions. During the 44–48 hour interval, three conditions are simultaneously satisfied: rapid hydrogen generation due to zirconium oxidation, residual oxygen at approximately 8--10\%, and a relatively low steam fraction. Their combination gives rise to a period of potential flammability. In contrast, during the 50–72 hour interval following the ECCS valve opening, oxygen dilution by the blowdown and the enrichment of hydrogen far beyond the stoichiometric proportion dominate, maintaining non-flammable conditions despite radiolytic oxygen accumulation. These results demonstrate that hydrogen flammability within the containment is not determined simply by a monotonic increase in oxygen concentration, but rather by the time-dependent interplay among the hydrogen generation and release history, the oxygen concentration, and the steam fraction. In other words, flammability is governed by the trajectory of the mixture composition relative to the flammable envelope. Notably, within the rich-hydrogen regime assessed here, the interval of greatest flammability risk is not the 72-hour time point at which the radiolytic oxygen criterion is approached, but instead the transient period (44–48 hours) during which vigorous hydrogen generation occurs while the mixture has not yet become deeply over-rich. This underscores the necessity of a thermodynamic-based flammability assessment that comprehensively accounts for time-dependent gas composition evolution.

\begin{table}[h]
\centering
\caption{Time-dependent gas composition, pressure, calculated adiabatic flame temperature, equivalence ratio, and flammability assessment.} 
\label{tab:flammability_results}
\resizebox{\textwidth}{!}{
\begin{tabular}{|c|c|c|c|c|c|c|c|c|}
\hline
Hour & $H_2O$(\%) & $H_2$(\%) & $O_2$(\%) & $N_2$(\%) & P (Pa) & \begin{tabular}[c]{@{}c@{}}Calculated Flame \\ Temperature (K)\end{tabular} & \begin{tabular}[c]{@{}c@{}}Equivalence \\ Ratio $\phi$ (--)\end{tabular} & \begin{tabular}[c]{@{}c@{}}Flammability\\  Assessment\end{tabular} \\ \hline
44   & 31.42      & 44.98     & 10.43     & 13.63     & 113044 & 1834.52                                                                     & 2.16                                                         & O                                                                  \\ \hline
46   & 26.13      & 55.20     & 8.25      & 10.52     & 161535 & 1552.52                                                                     & 3.35                                                         & O                                                                  \\ \hline
48   & 25.38      & 56.54     & 8.11      & 10.01     & 168244 & 1536.48                                                                     & 3.49                                                         & O                                                                  \\ \hline
50   & 24.41      & 66.30     & 4.44      & 5.04      & 300857 & 1006.03                                                                     & 7.47                                                         & X                                                                  \\ \hline
54   & 31.19      & 60.17     & 4.11      & 4.58      & 343244 & 943.93                                                                      & 7.32                                                         & X                                                                  \\ \hline
60   & 29.01      & 62.02     & 4.34      & 4.76      & 333842 & 981.62                                                                      & 7.15                                                         & X                                                                  \\ \hline
66   & 27.98      & 62.77     & 4.58      & 4.76      & 330569 & 1017.81                                                                     & 6.85                                                         & X                                                                  \\ \hline
72   & 27.37      & 63.28     & 4.67      & 4.86      & 328967 & 1032.57                                                                     & 6.78                                                         & X                                                                  \\ \hline
\end{tabular}
} 
\end{table}

\section{Discussion}

The results of this study carry the following implications for hydrogen safety in SMRs.

First, the necessity of flammability assessment under rich-hydrogen conditions has been confirmed. While lean-hydrogen conditions have been the primary concern in conventional large-scale NPPs, the limited containment volume of SMRs can give rise to rich-hydrogen conditions. In the LCC-05T-03 scenario, the containment atmosphere crosses stoichiometry at approximately 43 hours and remains fuel-rich thereafter. The hydrogen fraction reaches 45--66\% and the equivalence ratio ranges from about 2 to 7 (Table~\ref{tab:flammability_results}). These states lie far outside the lean regime addressed by existing hydrogen-specific models such as CNAFT, so a dedicated rich-side criterion is required. The CAFT-based criterion captures every flammable state represented by the FITS boundary across steam fractions of 5--50\%. It can serve as a viable screening tool for flammability assessment under rich-hydrogen conditions.

Second, a mechanistic rationale for a fixed thermal criterion on the rich side has been established. In rich hydrogen flames the deficient reactant is oxygen. Its mixture-averaged Lewis number is 1.35--2.10 along the FITS boundary and approximately 1.9--2.0 for the accident compositions. The preferential-diffusion amplification of the local flame temperature, expressed by Eq.~\ref{eq:caft_flameball}, is therefore absent. A Lewis number above unity also tends to suppress diffusional--thermal cellular instability, and the burned-gas temperature stays close to the CAFT. The FITS validation quantifies the resulting constancy, with the limit temperature varying within 17\% over the full steam range. The residual variation grows systematically with steam fraction, which is consistent with the radiative heat loss of steam.

Third, a flammability period may exist during a specific time interval of the accident progression. The flammable window during the 44--48 hour interval results from a combination of three conditions. The oxygen concentration remains at 8--10\% while hydrogen generation is vigorous. The steam fraction stays low, and the mixture has not yet become strongly over-rich. Follow-up analyses of this interval should therefore address three aspects. CFD analyses can resolve local mixing and stratification within the window. Screening for flame acceleration and deflagration-to-detonation transition can bound the consequences of ignition during this period. An inventory of credible ignition sources can establish the likelihood of combustion.

Fourth, the oxygen concentration criterion alone is insufficient for a comprehensive evaluation of hydrogen combustion risk in SMRs. The two approaches differ first in the timing of the assessment. The oxygen criterion anchors the evaluation to a single bounding state at 72 hours. The CAFT criterion resolves the transient and thereby uncovers the 44--48 hour window. The approaches differ also in the form of the result. The CAFT criterion yields a continuous margin, which falls from about $+722$ K at 44 hours to about $-80$ K at 72 hours. A concentration criterion returns only a binary verdict at each time. Finally, the thermodynamic criterion extends naturally to non-air backgrounds. The 72-hour mixture reaches an equivalence ratio of about 6.8 with only about 5\% nitrogen. Such states lie far outside the composition range of air-based flammability charts. The application of thermodynamic assessment methods that account for both mixture composition and the thermal effect of inert gases is therefore recommended for SMR containments.

However, the present study has the following limitations that require further investigation.

First, there is a limitation arising from the adiabatic assumption of the CAFT model. As discussed in Section 3.3, the conservatism of the criterion grows with steam fraction, a trend consistent with the radiative heat loss of steam. The development of a non-adiabatic correction model that accounts for the radiative heat loss effects of steam under rich-hydrogen conditions remains a subject for future work. Whether the CNAFT model approach developed by Jeon et al. for lean hydrogen can be extended to rich-hydrogen conditions merits further examination \cite{jeon2021mechanistic}. The sensitivity of the critical temperature to pressure and to third-body collision partners under rich conditions likewise merits quantification.

Second, the present study assumes that the gas composition within the containment vessel is fully premixed. Under actual severe accident conditions, gas mixing behavior may be influenced by buoyancy-driven flows and stratification. Therefore, CFD-based analyses of gas mixing behavior should be conducted in parallel to improve the fidelity of the flammability assessment. The present assessment is also a post-processing screening along the no-combustion MELCOR trajectory. An actual ignition would alter the subsequent pressure and composition evolution.

Third, the quantitative assessment in Section 4 was confined to the rich-hydrogen phase (after approximately 44 hours), for which the CAFT framework was validated in Section 3. The preceding hydrogen-lean phase begins at approximately 22–24 hours, when the steam fraction falls below the empirical inerting level. It ends at the stoichiometric crossover at approximately 43 hours. During this phase the containment atmosphere already satisfies conventional composition-based flammability criteria. The rich-hydrogen flammable window identified in this study should therefore be regarded as a lower bound on the duration of potentially flammable conditions over the transient. A quantitative evaluation of the lean phase using the non-adiabatic CNAFT model developed for lean hydrogen mixtures  is identified as future work \cite{jeon2019flammability, jeon2021mechanistic}.

Fourth, the threshold was derived from the fitted FITS boundary rather than from the individual test outcomes. A point-by-point evaluation against the raw burn and no-burn results of the 239 tests would provide an independent classification test. This evaluation is left for future work.

Fifth, the FITS boundary retains an air-derived nitrogen-to-oxygen ratio of approximately 3.76. The late-phase accident mixtures reach a ratio near unity because radiolysis adds oxygen while nitrogen is diluted. Applying the threshold to such states is a thermodynamically motivated extrapolation beyond the experimentally covered background compositions. The 400 K margin in the flammable window and the ERI AICC assessment at 72 hours mitigate the practical impact of this extrapolation.

\section{Conclusions}

In this study, the applicability of the CAFT methodology under rich-hydrogen conditions was examined mechanistically and validated empirically against the FITS experimental database. The validated criterion was then applied to the NuScale severe accident scenario LCC-05T-03 to evaluate hydrogen flammability within the containment vessel. The principal conclusions are as follows.

First, the CAFT evaluated along the measured FITS rich-limit boundary remains quasi-constant within 17\% across steam fractions of 5--50\%. Its minimum value of 1112.5 K serves as a conservative flammability threshold. This criterion captures every flammable state represented by the measured boundary, and misclassification is confined to the conservative side. The threshold also agrees within 5\% with independently reported critical flame temperatures of 1140--1168 K \cite{kumar1985flammability, terpstra2012flammability}. These results demonstrate that the CAFT methodology is a viable analytical tool for flammability assessment under rich-hydrogen conditions.

Second, the analysis of the NuScale SMR severe accident scenario identified a potential rich-hydrogen flammability period during the 44–48 hour interval following accident initiation. During this interval, the hydrogen fraction increases sharply due to zirconium oxidation, while the oxygen concentration is maintained at approximately 8–10\%. Furthermore, the relatively low steam fraction limits the steam inerting effect. Under these conditions, the adiabatic flame temperature is predicted to exceed 1500 K. This surpasses the threshold of 1112.5 K by more than 400 K. The preceding hydrogen-lean phase (approximately 22–43 hours) also satisfies conventional composition-based flammability criteria. The 44--48 hour window therefore represents a lower bound on the duration of potentially flammable conditions.

Third, the present results complement the bounding assessment framework documented in the ERI report. In that framework, hydrogen combustion was intentionally suppressed prior to 72 hours. The 72-hour state was adopted as the bounding evaluation point based on the oxygen concentration criterion. At that state, radiolytic oxygen accumulation brings the containment oxygen mole fraction to approximately 0.047. The CAFT analysis in this study identifies a transient flammable window during the 44–48 hour interval that lies outside the scope of that framework. It simultaneously corroborates, from a thermodynamic perspective, the ERI observation that the 72-hour mixture does not support combustion. This difference in coverage arises because the oxygen concentration criterion considers only the availability of oxidizer. The CAFT methodology accounts for both the position of the mixture relative to the flammable envelope and the overall heat capacity of the mixture. The former includes the over-rich state in which the hydrogen fraction exceeds the UFL. The gas composition changes rapidly during transient phases such as the blowdown following ECCS actuation. A thermodynamic framework therefore provides a more comprehensive flammability evaluation in such phases.

In reactor designs such as SMRs, the limited containment volume increases the likelihood of rich-hydrogen conditions. The application of thermodynamic flammability assessment methods can therefore enhance the reliability of hydrogen combustion risk evaluation. Future work should proceed in three directions. CFD analyses of gas mixing behavior can test the premixed assumption. A non-adiabatic correction model can reduce the conservatism under high-steam conditions. The CNAFT approach can extend the time-dependent assessment to the hydrogen-lean phase of the transient.

\appendix

\setcounter{table}{0}
\setcounter{figure}{0}

\section{FITS upper flammability limit boundary data used for CAFT validation}
\label{app:fits_data}

Table~\ref{tab:fits_boundary} lists the 21 mixture compositions used for the
CAFT validation in Section~3. The experimental UFL boundary of the
H$_2$--air--steam flammability diagram reported by Marshall~\cite{marshall1986hydrogen}
was discretized at uniform steam-fraction intervals
($\Delta x_{\mathrm{H_2O}} \approx 2.25\%$, spanning approximately 5--50\%),
and the corresponding hydrogen, oxygen, and nitrogen fractions were read from
the boundary curve. For all points, the initial temperature and air partial
pressure are 383~K and 83~kPa, respectively, consistent with the FITS
experimental baseline.

\begin{table}[htbp]
\centering
\caption{Mixture compositions discretized at uniform steam-fraction intervals
along the experimental UFL boundary of the FITS database~\cite{marshall1986hydrogen},
together with the calculated limit flame temperatures. Compositions are given in mol\%.
$\Delta T = T_{\mathrm{CAFT}} - T_{\mathrm{Threshold}}$ is the deviation from the data-derived
threshold $T_{\mathrm{Threshold}} = 1112.5$~K and $\delta = \Delta T / T_{\mathrm{Threshold}}$ its
relative value. All deviations are non-negative because the threshold is defined as the
minimum of the dataset.}
\label{tab:fits_boundary}
\small
\setlength{\tabcolsep}{4.5pt}
\begin{tabular}{@{}ccccccrrr@{}}
\toprule
No. & $x_{\mathrm{H_2O}}$ & $x_{\mathrm{H_2}}$ & $x_{\mathrm{O_2}}$ & $x_{\mathrm{N_2}}$ &
$\phi$ & \multicolumn{1}{c}{$T_{\mathrm{CAFT}}$ (K)} & \multicolumn{1}{c}{$\Delta T$ (K)} &
\multicolumn{1}{c}{$\delta$ (\%)} \\
\midrule
1 & 5.01 & 72.42 & 4.74 & 17.83 & 7.64 & 1112.5 & $0.0$ & $0.0$ \\
2 & 7.25 & 69.78 & 4.82 & 18.15 & 7.23 & 1118.4 & $5.9$ & $0.5$ \\
3 & 9.49 & 67.14 & 4.91 & 18.47 & 6.84 & 1124.2 & $11.7$ & $1.1$ \\
4 & 11.77 & 64.37 & 5.01 & 18.85 & 6.42 & 1132.3 & $19.8$ & $1.8$ \\
5 & 14.06 & 61.73 & 5.08 & 19.13 & 6.07 & 1136.2 & $23.7$ & $2.1$ \\
6 & 16.25 & 59.09 & 5.18 & 19.48 & 5.70 & 1143.3 & $30.8$ & $2.8$ \\
7 & 18.48 & 56.32 & 5.29 & 19.90 & 5.32 & 1152.7 & $40.2$ & $3.6$ \\
8 & 20.77 & 53.48 & 5.41 & 20.34 & 4.95 & 1162.3 & $49.8$ & $4.5$ \\
9 & 23.01 & 50.78 & 5.50 & 20.71 & 4.61 & 1169.4 & $56.9$ & $5.1$ \\
10 & 25.25 & 47.88 & 5.64 & 21.23 & 4.24 & 1182.2 & $69.7$ & $6.3$ \\
11 & 27.48 & 45.17 & 5.74 & 21.60 & 3.93 & 1189.0 & $76.5$ & $6.9$ \\
12 & 29.77 & 42.40 & 5.84 & 21.98 & 3.63 & 1196.1 & $83.6$ & $7.5$ \\
13 & 32.01 & 39.50 & 5.98 & 22.51 & 3.30 & 1208.4 & $95.9$ & $8.6$ \\
14 & 34.25 & 36.66 & 6.11 & 22.98 & 3.00 & 1218.6 & $106.1$ & $9.5$ \\
15 & 36.48 & 33.83 & 6.23 & 23.45 & 2.71 & 1228.7 & $116.2$ & $10.4$ \\
16 & 38.77 & 30.86 & 6.38 & 23.99 & 2.42 & 1240.7 & $128.2$ & $11.5$ \\
17 & 41.01 & 28.02 & 6.50 & 24.47 & 2.15 & 1250.4 & $137.9$ & $12.4$ \\
18 & 43.25 & 25.19 & 6.63 & 24.94 & 1.90 & 1260.0 & $147.5$ & $13.3$ \\
19 & 45.48 & 22.35 & 6.76 & 25.41 & 1.65 & 1269.4 & $156.9$ & $14.1$ \\
20 & 47.77 & 19.25 & 6.93 & 26.05 & 1.39 & 1284.4 & $171.9$ & $15.5$ \\
21 & 50.01 & 16.22 & 7.09 & 26.68 & 1.14 & 1298.8 & $186.3$ & $16.7$ \\
\bottomrule
\end{tabular}
\end{table}

\section*{Acknowledgment}

This work was supported by the Korea Institute of Energy Technology Evaluation and Planning (KETEP) and the Ministry of Climate, Energy $\&$ Environment (MCEE) of the Republic of Korea (No. RS-2024-00398425), and by the Nuclear Safety Research Program through the Korea Foundation of Nuclear Safety (KoFONS) using the financial resource granted by the Nuclear Safety and Security Commission (NSSC) of the Republic of Korea (No. RS-2026-25538145).

\bibliography{ref}
\bibliographystyle{unsrtnat}

\end{document}